\documentclass[11pt,letterpaper]{article}
\usepackage[margin=1in]{geometry}
\usepackage{amsmath,amssymb,amsthm}
\usepackage{graphicx,booktabs,placeins}
\usepackage{tikz}
\usetikzlibrary{arrows.meta}
\usepackage[round,authoryear]{natbib}
\usepackage{xurl}
\usepackage{microtype}
\usepackage[hidelinks]{hyperref}
\hypersetup{pdftitle={Conformal prediction for censored life history data},pdfauthor={Lu Mao}}
\newtheorem{theorem}{Theorem}
\newtheorem{proposition}{Proposition}
\newtheorem{corollary}{Corollary}
\theoremstyle{definition}
\newtheorem{definition}{Definition}
\newcommand{\pr}{\operatorname{pr}}
\newcommand{\E}{E}
\newcommand{\var}{\operatorname{var}}
\newcommand{\ind}{I}
\newcommand{\R}{\mathbb R}
\newcommand{\dd}{\,\mathrm{d}}
\newcommand{\calC}{\mathcal C}
\newcommand{\calD}{\mathcal D}

\title{Conformal prediction for censored life history data}
\author{Lu Mao\\[0.35em]
  \small Department of Biostatistics and Medical Informatics\\
  \small University of Wisconsin--Madison\\
  \small 610 Walnut Street, Madison, Wisconsin 53726, U.S.A.\\
  \small \href{mailto:lmao@biostat.wisc.edu}{\texttt{lmao@biostat.wisc.edu}}}
\date{}

\begin{document}
\maketitle

\begin{abstract}
We consider individual prediction of general summaries of a multistate trajectory, such as time alive out of hospital or time spent in remission. Right censoring complicates calibration because these outcomes may be only partially observed. We develop a conformal prediction approach for scalar and vector summaries over a fixed clinical horizon supported by follow-up. For each candidate region, we define the coverage determining time as the earliest time at which the observed history establishes whether the outcome lies in the region, even if the outcome itself is still unknown. We weight observed coverage decisions by the inverse probability of remaining uncensored to that time and augment the estimator with conditional coverage probabilities for unresolved histories. For a fixed region, the coverage estimator is doubly robust with respect to
the censoring and conditional coverage models and attains the semiparametric
efficiency bound when both are consistent at suitable rates. For the calibrated
region, we derive a finite-sample lower bound on the coverage probability 
and establish
asymptotic validity if either model is correct. In simulations, using partial histories yields narrower prediction intervals at comparable coverage. We illustrate joint prediction of days in and out of hospital in the HF-ACTION trial and prediction of relapse-free time in the German Breast Cancer Study Group cohort.
\end{abstract}


\section{Introduction}
At diagnosis, a clinician may want to tell a patient how long they are
likely to remain alive or how much time they may spend alive and out of
hospital over the next year. For cancer patients, time spent in relapse may also inform
treatment and supportive care. These prediction targets can be viewed as functionals
of a multistate process describing the patient's movement among health
states, such as home, hospital and death, or remission, relapse, metastasis
and death. The resulting life history then contains the patient's cumulative
experience, including the states visited and the time spent in each
\citep{andersen2002,putter2007,cook2018}.

Multistate models are natural tools for modelling such data. A multistate
model specifies transition intensities, the instantaneous rates of movement
between states, as functions of event history. Under the Markov assumption,
for example, these rates depend on the history only through the current
state and baseline characteristics. The fitted intensities can then be used to obtain the mean time spent in
selected states, or mean occupation time, for patients with given
characteristics \citep{buhler2023,buhler2025}. However, these population means
do not capture how occupation times vary across individuals. Prediction
intervals address this variation, but intervals obtained directly from a fitted outcome model have coverage that generally depends on its distributional assumptions. These assumptions can be particularly difficult to justify for complex life histories.

Conformal prediction provides a nonparametric approach to individual prediction 
through empirical calibration \citep{shafer2008,lei2018,angelopoulos2026foundations}. In the split-sample approach, an independent calibration sample estimates how often a fitted model's proposed regions cover the outcomes and determines their size to achieve the desired coverage rate. 
With fully observed outcomes, exchangeability of the calibration patients
and the new patient suffices to guarantee finite-sample marginal coverage
of at least the nominal level. This coverage is averaged over both calibration
samples and new patients and does not require the prediction model to be
correct.

For survival outcomes, censoring may end follow-up before the event, preventing some coverage checks during calibration. Under type-I censoring,
each patient's potential censoring time is known, as when follow-up ends at a
prespecified study date. Patients with sufficient potential follow-up can then
be used to calibrate lower prediction bounds by weighting
\citep{candes2023,gui2024}. Under ordinary right censoring, \citet{sesia2025}
extend this approach by imputing the potential censoring times hidden by
earlier events. \citet{farina2026} instead estimate coverage using inverse censoring
weights and augment the estimator with predictions of coverage based on
each patient's covariates and event-free follow-up. This augmentation
makes more efficient use of the calibration data and yields a consistent
estimator if either the censoring model or the survival model is correct.
Most methods for censored survival outcomes so far have focused on
lower prediction bounds. Informative upper bounds require follow-up long enough
to identify the relevant part of the upper tail \citep{farina2026}.

Extending univariate survival calibration to multistate outcomes requires
accounting for a richer outcome history. For a single event time, the
pre-event history records only how long the patient has remained
event-free. A multistate history also records transitions, such as
hospital admissions and discharges, and the time spent in each state.
This history can determine coverage before the final outcome is known.
For example, if a patient's hospital days to date exceed the upper limit
of a proposed interval, the total at the prediction horizon must also
exceed that limit. Similarly, when coverage remains uncertain, augmentation requires
estimating the probability of eventual coverage given the patient's
history to date.

We propose a calibration framework for general summaries of multistate
outcomes. By restricting the prediction target to a fixed clinical horizon
supported by follow-up, we can identify both lower and upper prediction
bounds under independent censoring. Predicting restricted occupation
times for individual patients complements inference on the corresponding
population means \citep{buhler2023,buhler2025}, including the restricted
mean survival time \citep{uno2014}. 

As follow-up unfolds, we track the target values
compatible with the patient's observed history. The first time this
feasible set lies entirely inside or outside a proposed region
is the coverage determining time. We can use this decision even
when the final outcome is unknown, weighting it by the inverse
probability of remaining uncensored to that time. We augment the
weighted estimator with predictions of coverage for unresolved cases,
using either Markov transition calculations that incorporate accrued
occupation times or simulations of the remaining trajectory. For a
fixed region, the augmented estimator is doubly robust and attains the
semiparametric efficiency bound when the censoring and conditional coverage estimators
converge at suitable rates. For the calibrated region, we derive a
finite-sample lower bound on coverage and show that asymptotic coverage
is at least the nominal level if either model is correct. The R package
\texttt{lifeconformal} implements these methods and is publicly
available at
\url{https://github.com/lmaowisc/lifeconformal}.

\section{Prediction targets and observed histories}\label{sec:framework}
\subsection{Targets as functionals of a full trajectory}\label{sec:targets}
\begin{figure}[ht]
\centering

\tikzset{working state/.style={rectangle,draw,thick,
  minimum width=1.5cm,minimum height=0.9cm,align=center}}
\begin{minipage}[t]{0.49\textwidth}
\centering
\scalebox{0.8}{(a) Repeated hospitalisation}\par\smallskip
\begin{tikzpicture}[x=\linewidth,y=1cm,>=Stealth,
  every node/.style={font=\normalsize,scale=0.85}]
  \path[use as bounding box] (0,-2.25) rectangle (1,0.90);
  \node[working state] (h0) at (0.15,0) {0: Home};
  \node[working state] (h1) at (0.85,0) {1: Hospital};
  \node[working state] (h2) at (0.50,-1.8) {2: Death};
  \draw[->,thick] ([yshift=3pt]h0.east) -- ([yshift=3pt]h1.west);
  \draw[->,thick] ([yshift=-3pt]h1.west) -- ([yshift=-3pt]h0.east);
  \draw[->,thick] (h0) -- (h2);
  \draw[->,thick] (h1) -- (h2);
\end{tikzpicture}
\end{minipage}\hfill
\begin{minipage}[t]{0.49\textwidth}
\centering
\scalebox{0.8}{(b) Cancer patients}\par\smallskip
\begin{tikzpicture}[x=\linewidth,y=1cm,>=Stealth,
  every node/.style={font=\normalsize,scale=0.85}]
  \path[use as bounding box] (0,-2.25) rectangle (1,0.90);
  \node[working state] (o0) at (0.15,0) {0: Remission};
  \node[working state] (o1) at (0.50,0) {1: Relapse};
  \node[working state] (o2) at (0.85,0) {2: Metastasis};
  \node[working state] (o3) at (0.50,-1.8) {3: Death};
  \draw[->,thick] (o0) -- (o1);
  \draw[->,thick] (o1) -- (o2);
  \draw[->,thick,rounded corners=2pt] (o0.north) -- ++(0,0.28) -| (o2.north);
  \draw[->,thick] (o0) -- (o3);
  \draw[->,thick] (o1) -- (o3);
  \draw[->,thick] (o2) -- (o3);
\end{tikzpicture}
\end{minipage}
\caption{Two examples of multistate models.}
\label{fig:state-models}
\end{figure}
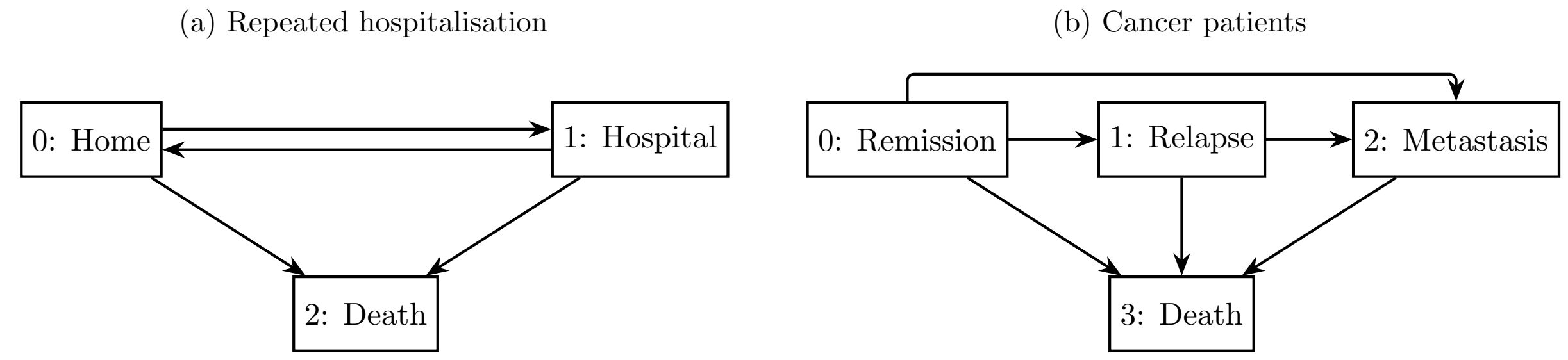

Let $Y(t)\in\mathcal S=\{0,1,\ldots,K\}$ be a right-continuous
multistate process and $\mathcal H(t)=\{Y(u):0\le u\le t\}$ the
life history through time $t$. The state labels are arbitrary and
unordered. A multistate model specifies the distinct states and
possible transitions between them. Fig.~\ref{fig:state-models}
illustrates two simple examples in diagrams: repeated hospital admission
and discharge before
death and progression from remission to relapse, metastasis or death.
In both examples, death is the only absorbing state, that is, a state
that cannot be left once entered. More generally, $\mathcal S$ may
contain any number of absorbing states, including none. By definition, 
entry into
an absorbing state fixes the subsequent trajectory.

For a fixed horizon $\tau>0$, let $\Gamma(\mathcal H)(\tau)\in\mathcal R\subseteq\R^d$ be a scalar or vector functional of the trajectory through $\tau$, with shorthand $\Gamma(\tau)$ when the trajectory is understood. 
Our principal targets are state-occupation times $\Gamma(\tau)=\{\Gamma_{\mathcal A_1}(\tau),\ldots, \Gamma_{\mathcal A_d}(\tau)\}^{\mathrm T}$, where
\[
\Gamma_{\mathcal A}(\tau)=\int_0^\tau I\{Y(t)\in\mathcal A\}\dd t,
 \quad \mathcal A\subseteq\mathcal S,
\]
and $I(\cdot)$ denotes the indicator function. In the hospitalisation model, $\Gamma_{\{0\}}(\tau)$ is time alive out of hospital and $\Gamma_{\{1\}}(\tau)$ is the total time spent in hospital. In the cancer model, $\Gamma_{\{1\}}(\tau)$ records time in relapse, while $\Gamma_{\{1,2\}}(\tau)$ records time in relapse or metastasis. In both examples, taking $\mathcal A$ to be all living states gives the restricted survival time, $\Gamma_{\mathcal A}(\tau)=D\wedge\tau$, where $D$ denotes time to death and $a\wedge b=\min(a,b)$. Its expectation is the restricted mean survival time \citep{uno2014}.

Besides occupation times, other trajectory functionals can summarise the severity of a patient's course. When the state labels increase with severity, for example, $\Gamma(\tau)=\max_{0\le t\le\tau}Y(t)$ records the most severe state reached by $\tau$. In Fig.~\ref{fig:state-models}(a), its values $0,1,2$ indicate being alive without hospitalisation, alive after at least one hospitalisation, or dead at $\tau$, respectively. In Fig.~\ref{fig:state-models}(b), the corresponding categories are continued remission, relapse without metastasis or death, metastasis while alive, and death.

Let $Z$ contain the baseline characteristics used for prediction, including the initial state $Y(0)$ if it varies, and let $\mathcal R(Z)\subseteq\mathcal R$ be the target range compatible with this baseline information. For a new patient with covariates $Z^*$ and a prescribed miscoverage level $\alpha\in(0,1)$, we seek a region containing the entire target vector with probability at least $1-\alpha$, up to an error that vanishes with sample size. This coverage probability is marginal, averaging over the data used to
construct the region and over the new patient's baseline covariates and
trajectory \citep[see, e.g., Section~3.2 of ][]{angelopoulos2026foundations}.

In the split-sample approach, we fit a working outcome model, such as
a multistate Markov model, on an independent training sample
$\calD_{\rm tr}$. We then define a score function $s(z,\gamma)$ to
measure the nonconformity of a hypothetical target value $\gamma$
with the model prediction for covariates $z$. A common score for
real-valued targets is
$s(z,\gamma)=|\gamma-\widehat\mu(z)|/\widehat\sigma(z)$, which measures
departure from the model-predicted mean $\widehat\mu(z)$ in units of
the model-predicted standard deviation $\widehat\sigma(z)$.
Score choices and model-fitting details are discussed in
Section~\ref{sec:implementation}.

In general, a prediction region includes target values whose
nonconformity scores do not exceed a threshold \(q\):
\[
 \calC_q(z)=\{\gamma\in\mathcal R(z):s(z,\gamma)\le q\}.
\]
To achieve a desired coverage probability, we could choose $q$ 
as a quantile of the score distribution implied by the fitted outcome distribution. However, this would make coverage dependent on the correctness
of the working model. For robustness, we choose
$q$ using an independent calibration sample, with
the trained score function held fixed. The calibration sample may
also be censored, so we first examine how censoring affects our
ability to discern possible target values.

\subsection{Feasible sets during follow-up}\label{sec:determination}
If a patient dies before $\tau$, their time alive out of hospital through $\tau$ is already known. Time in relapse can be known even before death: in Fig.~\ref{fig:state-models}(b), entry into metastasis fixes the time spent in relapse because there is no path back to relapse. More generally, when $Y(t)\notin\mathcal A$ and no path leads from $Y(t)$ to $\mathcal A$, the accrued time $\Gamma_{\mathcal A}(t)$ is also the final value $\Gamma_{\mathcal A}(\tau)$.

Even when the target remains unknown, the observed history narrows its possible values
as follow-up goes on. For occupation time in $\mathcal A$, the accrued time $\Gamma_{\mathcal A}(t)$ gives a lower bound. At most $\tau-t$ can be added, giving the outer interval $[\Gamma_{\mathcal A}(t),\ \Gamma_{\mathcal A}(t)+\tau-t]$.
This interval starts at $[0,\tau]$ and shrinks as time is spent inside or outside $\mathcal A$. Transition restrictions may narrow it further, as in the relapse example above.
The same reasoning applies to the most severe state reached by $\tau$. In Fig.~\ref{fig:state-models}(a), a hospital admission rules out $0$, leaving the possible values $1$ and $2$; death fixes the value at $2$.

For a general target, we can define its feasible set at time $t$
in the following way. Let the trajectory vary over all admissible paths through $\tau$, with its segment over $[0,t]$ tied down to the observed history. Applying the target functional to these completed trajectories gives a range of possible values, which we call the feasible set. Let $\mathbb H_\tau(Z)$ be the class of trajectories through $\tau$ obeying the known transition and baseline constraints. The feasible
set for $\Gamma(\mathcal H)(\tau)$ at $t$ is
\begin{equation*}
 \mathcal K_\Gamma(\mathcal H)(t;\tau)
 =\{\Gamma(\mathcal H^*)(\tau)\in\mathcal R(Z):
       \mathcal H^*(t)=\mathcal H(t),\ 
       \mathcal H^*(\tau)\in\mathbb H_\tau(Z)\}.
\end{equation*}
Write $\mathcal K_\Gamma(t;\tau)$ when that history is understood. A longer history leaves fewer continuations, so these sets are nested: for $0\le t_1\le t_2\le\tau$,
\begin{equation*}
 \{\Gamma(\tau)\}=\mathcal K_\Gamma(\tau;\tau)
 \subseteq\mathcal K_\Gamma(t_2;\tau)
 \subseteq\mathcal K_\Gamma(t_1;\tau)
 \subseteq\mathcal K_\Gamma(0;\tau)=\mathcal R(Z).
\end{equation*}
Thus, the feasible set shrinks from the range allowed at baseline to the target singleton. 
As a result, to determine whether $\Gamma(\tau)\in\mathcal C_q(Z)$, we only
need enough follow-up so that the feasible set lies entirely inside or outside
$\calC_q(Z)$.

\begin{figure}[ht]
\centering

\includegraphics[width=\textwidth]{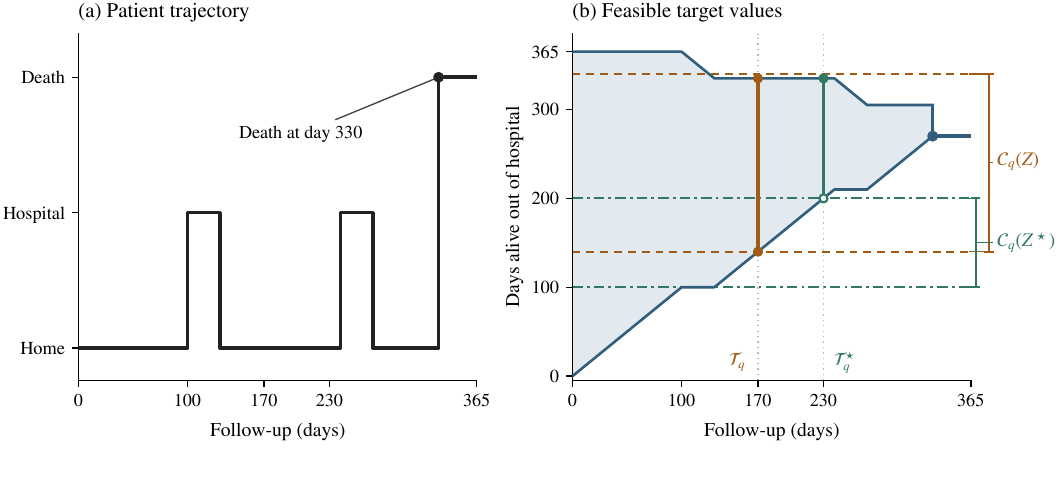}
\caption[Patient trajectory and shrinking feasible target interval]{Patient trajectory (left) and feasible values of days alive out of hospital over 365 days (right). The shaded area represents feasible target values.}
\label{fig:feasible-history}
\end{figure}

Fig.~\ref{fig:feasible-history} illustrates this concept for days alive out of hospital over 365 days. Spending time at home raises the lower bound, while time in hospital lowers the upper bound. Death on day 330 fixes the target at 270 days. Apply two hypothetical intervals, $\calC_q(Z)=[140,340]$ and $\calC_q(Z^\star)=[100,200]$, to this trajectory, and we find that coverage or non-coverage is determined on days 170 and 230, respectively, before the target is known.

Formally, we define the coverage determining time as the earliest time
coverage of $\Gamma(\tau)$ by $\calC_q(Z)$ is resolved.
\begin{definition}\label{def:coverage-time}
The coverage determining time is
\begin{equation*}
\begin{split}
 \mathcal T_{\Gamma,q}(\tau)=\inf\{t\in[0,\tau]:\quad
 &\mathcal K_\Gamma(t;\tau)\subseteq\calC_q(Z)\quad\text{or } \quad
 \mathcal K_\Gamma(t;\tau)\cap\calC_q(Z)=\varnothing\}.
\end{split}
\end{equation*}
Containment gives $\Gamma(\tau) \in \calC_q(Z)$; disjointness gives $\Gamma(\tau) \notin \calC_q(Z)$. For fixed $\Gamma$ and $\tau$, write $\mathcal T_q=\mathcal T_{\Gamma,q}(\tau)$.
\end{definition}

Since entry into an absorbing state fixes the subsequent trajectory,
we have
\begin{equation}\label{eq:ineq_cdt}
 \mathcal T_{\Gamma,q}(\tau) \le\tau\wedge D,
\end{equation}
where, with a slight abuse of notation, we use $D$ to denote the
entry time into any absorbing state in $\mathcal S$, with $D=+\infty$ if none is entered.

\section{Calibration from censored histories}\label{sec:calibration}
\subsection{Observation under censoring}
The working outcome model and score function have been fitted on the independent training sample. We hold them fixed throughout calibration and suppress conditioning on that sample.
For a generic patient, let $C$ denote the censoring time. We assume conditional independent censoring,
\begin{equation}\label{eq:cind}
 C\ \perp\ \mathcal H(t)\mid Z,\qquad t\ge0.
\end{equation}

 All transitions are observed through $X=C\wedge D$. With $\Delta=\ind(D\le C)$ indicating observed entry into an absorbing state, the observed data are $O=\{Z,X,\Delta,\mathcal H(X)\}$.
Given a random calibration sample $O_1,\ldots,O_n$ of independent copies of $O$, we estimate the coverage probability of $\calC_q$ across thresholds $q$. These estimates allow us to choose a threshold $\widehat q$ at which estimated coverage reaches the desired level, say, $1-\alpha= 90\%$. Censoring before $\tau$ leaves some target values unknown. Standard inverse weighting uses $\ind(C\ge\tau\wedge D)$ as the observation indicator, since follow-up through $\tau\wedge D$ determines the trajectory through $\tau$. For coverage estimation, however, we only need to know whether the target belongs to $\calC_q(Z)$, not its precise value. 
Recall from Section \ref{sec:determination} that this membership is resolved at $\mathcal T_q\le\tau\wedge D$.

We therefore use $R_q=\ind(C\ge\mathcal T_q)$ as the observation indicator for $\ind\{\Gamma(\tau)\in\calC_q(Z)\}$, with $X_q=C\wedge\mathcal T_q$ the follow-up available for determining coverage. Since $\mathcal T_q\le D$, both $R_q$ and $X_q$ are observed even when absorption hides $C$, as when a patient dies before censoring.

\subsection{Estimating coverage by inverse weighting}
To estimate coverage, write $\Gamma_\tau=\Gamma(\tau)$ for the full
target and define the coverage curve at each threshold $q$ by
\[
 F(q)=\pr\{\Gamma_\tau\in\calC_q(Z)\}
     =\pr\{s(Z,\Gamma_\tau)\le q\}.
\]
Write $G_0(t\mid Z)=\pr(C\ge t\mid Z)$ for the conditional censoring survival function and require $G_0(\tau\mid Z)\ge c_0>0$. Under~\eqref{eq:cind}, the probability that $R_q=1$, given $Z$ and the full trajectory, is $G_0(\mathcal T_q\mid Z)$. Weighting each observed coverage indicator by the inverse of this probability thus gives an unbiased contribution to $F(q)$.

We estimate $G_0$ by $\widehat G$ on the independent training sample; Section~\ref{sec:implementation} gives fitting details. We condition on the training data throughout calibration, so the fitted censoring weights are treated as fixed. Averaging the weighted coverage indicators over the calibration sample gives the inverse probability weighted (IPW) estimator
\begin{equation*}
 \widehat F_{\mathrm{IPW}}(q)=\frac1n\sum_{i=1}^n
       \frac{R_{qi}\ind\{\Gamma_{\tau,i}\in\calC_q(Z_i)\}}{\widehat G(X_{qi}\mid Z_i)}.
\end{equation*}
Here $i$ indexes the calibration patients.

Using coverage determining times earlier than absorption makes fuller use of the observed histories and can thus reduce variance. Indeed, let $\widehat F_{\mathrm{IPW},0}(q)$ denote the estimator above with $\widehat G=G_0$. Standard IPW uses $\Delta_\tau=\ind(C\geq \tau\wedge D)$ as its observation indicator, with $(\Delta_{\tau,i},X_i)$ denoting the corresponding sample copies of $(\Delta_\tau,X)$. Its estimator is $\widehat F_{\mathrm{IPW},0}^{\mathrm{std}}(q)=n^{-1}\sum_{i=1}^n\Delta_{\tau,i}\ind\{\Gamma_{\tau,i}\in\calC_q(Z_i)\}/G_0(X_i\wedge\tau\mid Z_i)$. For a fixed $q$, the variance difference is
\begin{align*}
 &\var\{\widehat F_{\mathrm{IPW},0}^{\mathrm{std}}(q)\}
 -\var\{\widehat F_{\mathrm{IPW},0}(q)\}\\
 &\qquad=n^{-1}\E\left[\ind\{\Gamma_\tau\in\calC_q(Z)\}\left\{\frac1{G_0(\tau\wedge D\mid Z)}-\frac1{G_0(\mathcal T_q\mid Z)}\right\}\right]\ge0.
\end{align*}
The difference is nonnegative because $\mathcal T_q\le\tau\wedge D$ by~\eqref{eq:ineq_cdt} and $G_0(t\mid Z)$ is nonincreasing in $t$. The gain can be substantial when covered trajectories are frequently censored between these times, since our estimator retains their known coverage indicators while standard IPW discards them. Section S2.6 of the Supplementary Material gives the proof and discusses estimated censoring weights.

\subsection{Efficient influence function}\label{sec:coverage-eif}
Histories censored before coverage determination do not resolve target membership 
but can still provide partial information. Such partial information is captured by the conditional coverage probability
\begin{equation*}
 Q_{0q}(t)=\pr\{\Gamma_\tau\in\calC_q(Z)\mid Z,\mathcal H(t-)\},\qquad t\le\mathcal T_q,
\end{equation*}
where $\mathcal H(t-)$ denotes the history strictly before $t$, including recorded transitions and accrued times. After determination, $Q_{0q}(t)=\ind\{\Gamma_\tau\in\calC_q(Z)\}$ for $\mathcal T_q<t\le\tau\wedge D$.

These conditional probabilities can improve the efficiency of the IPW estimator $\widehat F_{\mathrm{IPW}}(q)$. Formally, we derive the efficient influence function for $F(q)$ in the nonparametric observed-data model by perturbing the distributions of $Z$, the outcome trajectory and $C$, subject to~\eqref{eq:cind} \citep[Section~7.3]{tsiatis2006}. Write $\lambda_0(t\mid Z)$ for the true censoring hazard, so $G_0(t\mid Z)=\exp\{-\int_0^t\lambda_0(u\mid Z)\dd u\}$.

\begin{samepage}
\begin{theorem}[Efficient influence function]\label{thm:coverage-eif}
Hold the trained score $s(\cdot,\cdot)$ and threshold $q$ fixed. Under the
regularity conditions specified in
Section S2.1 of the Supplementary Material, the efficient influence
function for $F(q)$ is
\begin{equation}\label{eq:coverage-eif}
 \phi_q(O)
 =
 \frac{R_q\ind\{\Gamma_\tau\in\calC_q(Z)\}}
      {G_0(X_q\mid Z)}
 -F(q)
 +\int_0^{X_q}
   \frac{Q_{0q}(t)}{G_0(t\mid Z)}\dd M_{0q}^{C}(t),
\end{equation}
where
\[
 M_{0q}^{C}(t)
 =\ind\{X_q\le t,\ R_q=0\}
  -\int_0^{t\wedge X_q}\lambda_0(u\mid Z)\dd u
\]
is the censoring martingale stopped at coverage determination.
The efficiency bound per calibration observation is thus
$\E\{\phi_q(O)^2\}$.
\end{theorem}
\end{samepage}

The martingale term in~\eqref{eq:coverage-eif} uses partial histories to predict unresolved coverage, improving efficiency without changing the mean of the IPW contribution. Once coverage is determined, further follow-up adds no information about membership, so stopping at $\mathcal T_q$ preserves the efficient influence function and its bound. Section S2.5 of the Supplementary Material gives the derivation.

This construction extends the survival calibration approach of \citet{farina2026} and \citet{siqiu2026} to complex life history data 
and general trajectory targets. In univariate survival, the pre-event history records only survival to the current time. Here, $Q_{0q}(t)$ also conditions on the observed transitions and accrued times. Its calculation therefore requires the distribution of the target over possible continuations of that history.

In practice, the conditional coverage probabilities $Q_{0q}$ are unknown. We estimate them using a working outcome model fitted on the training data, possibly the same model used to train the score function. Substituting these estimates and the fitted censoring distribution into~\eqref{eq:coverage-eif} gives the augmented estimator below.

\subsection{Augmentation using partial histories}\label{sec:augmentation}
Let $\widehat{\pr}$ denote probability under the fitted outcome model.
For $t\le\mathcal T_q$, we estimate $Q_{0q}(t)$ by
\[
 \widehat m_q(t)
 =\widehat{\pr}\{\Gamma_\tau\in\calC_q(Z)\mid Z,\mathcal H(t-)\}.
\]
Section~\ref{sec:implementation} discusses models suited to multistate processes and dynamic prediction, along with their fitting procedures.

For generic functions $G$ and $m$, first suppose $G$ has a nonnegative hazard $\lambda$, so $G(t\mid Z)=\exp\{-\int_0^t\lambda(u\mid Z)\dd u\}$. By \eqref{eq:coverage-eif}, the corresponding augmented contribution is
\begin{equation*}
 \psi_q(O;G,m)=
 \frac{R_q\ind\{\Gamma_\tau\in\calC_q(Z)\}}
      {G(X_q\mid Z)}
 +\int_0^{X_q}
   \frac{m_q(t)}{G(t\mid Z)}\dd M_q^{C}(t),
\end{equation*}
where $M_q^{C}$ is obtained from $M_{0q}^{C}$ defined in Theorem \ref{thm:coverage-eif} by replacing $\lambda_0$ with the hazard $\lambda$ of $G$. This leads to the augmented inverse probability weighted (AIPW) estimator
\begin{equation}\label{eq:aipw}
 \widehat F_{\mathrm{AIPW}}(q)
 =\frac1n\sum_{i=1}^n\psi_q(O_i;\widehat G,\widehat m).
\end{equation}
At the true functions $G_0$ and $Q_0=\{Q_{0q}:q\in\R\}$, $\psi_q(O;G_0,Q_0)-F(q)=\phi_q(O)$. The augmented estimator is doubly robust with respect to $G$ and $m$ and attains the efficiency bound under the estimation conditions in Proposition~\ref{prop:coverage-efficiency} of Section~\ref{sec:theory}. Section~S2.8 of the Supplementary Material gives computational details.

To construct a prediction region with desired coverage rate $1-\alpha\in(0,1)$, we invert $\widehat F_{\mathrm{AIPW}}$ using the split conformal quantile rule \citep{angelopoulos2026foundations}. Specifically, set $k_n=\lceil(n+1)(1-\alpha)\rceil$ and choose the smallest threshold at which estimated coverage reaches $k_n/n$:
\begin{equation}\label{eq:cutoff}
 \widehat q=\inf\{q:\widehat F_{\mathrm{AIPW}}(q)\ge k_n/n\},
 \qquad\mbox{and }\quad \widehat{\calC}(z)=\calC_{\widehat q}(z).
\end{equation}
If $k_n=n+1$ or no finite threshold qualifies, we set $\widehat q=+\infty$ and return the full feasible space.

In the absence of censoring, setting $\widehat G\equiv1$ recovers ordinary split conformal prediction. Under censoring, the guarantees in Section~\ref{sec:theory} account for estimation error in $\widehat G$ and $\widehat m_q$. The estimated curve $\widehat F_{\mathrm{AIPW}}(\cdot)$ may optionally be monotonised and clipped to $[0,1]$ before inversion, as described in Section S4 of the Supplementary Material.

\subsection{Implementation details}\label{sec:implementation}
To implement the calibration in~\eqref{eq:cutoff}, we first need to specify the score
$s(z,\gamma)$. A standard choice for vector-valued targets is a weighted maximum of standardised component deviations. For occupation times, the fitted conditional means and variances are
\[
 \widehat\mu_j(z)=\widehat{\E}\{\Gamma_{\mathcal A_j}(\tau)\mid Z=z\},
 \qquad
 \widehat\sigma_j^2(z)=\widehat{\var}\{\Gamma_{\mathcal A_j}(\tau)\mid Z=z\}.
\]
Under a Markov working model, we can compute these moments backwards from $\tau$, where the remaining occupation time is zero, 
and use the fitted transition probabilities to update the moments at each step back to baseline. Alternatively, we can estimate them from occupation times simulated under the fitted outcome model. Sections S10.2--S10.3 of the Supplementary Material outline the moment recursions and simulation procedures.

We then standardise the componentwise deviations and combine them through a weighted maximum:
\begin{equation}\label{eq:maximum_joint}
    s_j(z,\gamma_j)=
 \frac{|\gamma_j-\widehat\mu_j(z)|}{\widehat\sigma_j(z)},
 \qquad
 s(z,\gamma)=\max_{1\le j\le d}\frac{s_j(z,\gamma_j)}{a_j}, 
\end{equation}
where $a_j>0$ control relative interval widths. In computing the scores, we bound each fitted standard deviation away from zero as described in Section~S10.1 of the Supplementary Material.
Quantile-based component scores can accommodate skewness \citep{romano2019,sampson2024}. All score fitting, including any tuning of flexible models, uses $\calD_{\rm tr}$, and the score is held fixed throughout calibration. A well-fitted score can yield smaller prediction regions, but its working model need not be correct for coverage.

Alternatively, we can calibrate each component separately at its own
nominal coverage level. Specifically, choose $\alpha_j>0$ in advance or using $\calD_{\rm tr}$, with $\sum_{j=1}^d\alpha_j\le\alpha$, assigning smaller $\alpha_j$ where higher coverage is desired and larger $\alpha_j$ where narrower intervals are preferred. For each component, we use the same $n$ calibration observations to estimate $\pr[s_j\{Z,\Gamma_{\mathcal A_j}(\tau)\}\le q]$ by~\eqref{eq:aipw}, using its scalar feasible sets and coverage determining times. This may allow us to determine component-specific coverage
even before joint coverage is resolved. We select $\widehat q_j$ by~\eqref{eq:cutoff} with $\alpha_j$ in place of $\alpha$. The component and joint prediction sets are
\begin{equation}\label{eq:componentwise_region}
     I_j(z)=\{\gamma_j:s_j(z,\gamma_j)\le\widehat q_j\},
 \qquad
 \widehat{\calC}_{\rm alloc}(z)
 =\mathcal R(z)\cap\prod_{j=1}^d I_j(z).
\end{equation}
Proposition~\ref{prop:unequal-allocation} gives the component-specific and joint coverage guarantees.
For coverage determination, Section S10.1 of the Supplementary Material gives an efficient algorithm for computing $\mathcal T_q$ jointly across thresholds.

In the split-sample procedure, we fit and tune the nuisance functions $\widehat G$ and $\widehat m$ in~\eqref{eq:aipw} on $\calD_{\rm tr}$. 
For $\widehat G$, a standard semiparametric model, such as the Cox model, typically
suffices.
To construct $\widehat m$, we can use simple Markov models or flexible alternatives that incorporate earlier history. The Markov approach is computationally simple and extends the earlier moment calculation to coverage probabilities. 
For coverage, we propagate the indicator of membership in the candidate region backwards from $\tau$ to $t$, then evaluate the resulting probability at the patient's current state and accrued occupation times to obtain $\widehat m_q(t)$. Section S10.2 of the Supplementary Material details this backward recursion.

Flexible transition models use more of the partial history and may improve the statistical efficiency of coverage estimation. For example, survival-tree ensembles or neural transition models \citep{rossman2022,cottin2026} can be used to estimate transition rates from baseline covariates, state durations, previous transition counts and accrued occupation times. Given $Z$ and $\mathcal H(t-)$, we simulate $B$ continuations by successively drawing transition times and destinations and updating the history through $\tau$. Combining each continuation with the observed history 
$\{Z,\mathcal H(t-)\}$ yields a completed target vector $\Gamma^{(b, t)}(\tau)$. Before coverage determination, the proportion of completed targets in the region approximates the fitted conditional coverage probability,
\[
 \begin{aligned}
 \widehat m_q(t)
 =\widehat{\pr}\{\Gamma_\tau\in\calC_q(Z)\mid Z,\mathcal H(t-)\}
 \approx B^{-1}\sum_{b=1}^B
 \ind\bigl\{\Gamma^{(b, t)}(\tau)\in\calC_q(Z)\bigr\}.
 \end{aligned}
\]
Section S10.3 of the Supplementary Material provides the fitting and simulation details.
 Under the conditions of Theorem~\ref{thm:identity} and Corollary~\ref{cor:asymptotic-coverage}, consistency of either $\widehat G$ or $\widehat m$ suffices for asymptotic coverage.

The calibration sample can also contribute to fitting $\widehat G$ and $\widehat m$. However, each observation should then be evaluated using functions fitted on other observations. We therefore divide the calibration sample into folds and cross-fit these functions, as in \citet{qiu2023}. For each fold, we fit them using $\calD_{\rm tr}$ and the remaining calibration folds, then compute the augmented contributions on the held-out fold. We pool these contributions before selecting the prediction threshold, keeping the trained score fixed throughout. Sections S7.1 and S10.1 of the Supplementary Material provide the construction and coverage guarantees.

\section{Theoretical guarantees}\label{sec:theory}
\subsection{Double robustness of coverage estimation}\label{sec:bias-efficiency}
The augmented estimator in~\eqref{eq:aipw} uses two fitted quantities: the censoring distribution and conditional coverage probabilities. Their errors interact only while coverage is unresolved. For generic $G$ and $m$, define
\[
 \|a\|_q^2=\E\left\{\int_0^{\mathcal T_q}a(t)^2\dd t\right\},\qquad
 r_C=\sup_q\|\lambda-\lambda_0\|_q,\quad
 r_Q=\sup_q\|m_q-Q_{0q}\|_q.
\]
The norm averages squared discrepancies over patients and time before determination. Thus, $r_C$ and $r_Q$ measure censoring-hazard and conditional-coverage errors uniformly over candidate thresholds. Recall that $c>0$ is a lower bound for $G(\tau\mid Z)$ under the regularity conditions in Section S2.1 of the Supplementary Material. 

\begin{theorem}[Bias identity]\label{thm:identity}
In the setting of Sections~\ref{sec:framework} and~\ref{sec:calibration}, and under the regularity conditions in Section S2.1 of the Supplementary Material,
\begin{align*}
 \E\left[\frac{R_q\ind\{\Gamma_\tau\in\calC_q(Z)\}}{G_0(X_q\mid Z)}\right]&=F(q),\\
 \E\{\psi_q(O;G,m)\}-F(q)
 &=\E\left[\int_0^{\mathcal T_q}\frac{G_0(t\mid Z)}{G(t\mid Z)}
 \{m_q(t)-Q_{0q}(t)\}\{\lambda_0(t\mid Z)-\lambda(t\mid Z)\}\dd t\right].
\end{align*}
Consequently, $\sup_q|\E(\psi_q)-F(q)|\le c^{-1}r_Cr_Q$.
\end{theorem}

The first identity justifies the inverse weighting. The second establishes double robustness: bias vanishes if either the censoring distribution or the conditional coverage probabilities are correct. As in \citet{farina2026}, the remaining bias is controlled by a product of nuisance errors. Here, however, the coverage error is assessed along partial multistate histories and only until $\mathcal T_q$. Intermediate states and accrued times can therefore reduce error both by settling coverage earlier and by improving conditional predictions. 

\subsection{Efficient estimation}
Theorem~\ref{thm:coverage-eif} identifies the efficient influence function. To attain its variance bound, the fitted augmented contributions must approach their true counterparts and their bias must be negligible at the $n^{-1/2}$ scale. Write $\widehat\lambda$ for the hazard of $\widehat G$ and $\psi_{0q}=\psi_q(\cdot;G_0,Q_0)$, let $P$ denote the observed-data law and define $\|f\|_{P,2}=\{\E f(O)^2\}^{1/2}$.

\begin{proposition}[Efficient estimation]\label{prop:coverage-efficiency}
Under the model and regularity conditions of Theorem~\ref{thm:coverage-eif}, hold $s$ and $q$ fixed. For nuisance functions fitted independently of the calibration sample, suppose that
\[
 \|\psi_q(\cdot;\widehat G,\widehat m)-\psi_{0q}\|_{P,2}=o_p(1),\qquad
 \|\widehat\lambda-\lambda_0\|_q\|\widehat m_q-Q_{0q}\|_q=o_p(n^{-1/2}).
\]
Then the augmented estimator satisfies
\begin{equation*}
 \sqrt n\{\widehat F_{\mathrm{AIPW}}(q)-F(q)\}
 =n^{-1/2}\sum_{i=1}^n\phi_q(O_i)+o_p(1),
\end{equation*}
with limiting variance $\E\{\phi_q(O)^2\}$, the efficiency bound per calibration observation, and is thus semiparametric efficient. 
\end{proposition}

The two conditions distinguish consistency from the rate needed for efficiency. Under suitable continuity of $\psi_q$, consistent estimation of both nuisance functions implies the first condition. The product-rate condition then makes the remaining bias negligible relative to $n^{-1/2}$ sampling variation. The estimator therefore behaves asymptotically as if the true nuisance functions were known and attains the efficiency bound. Section S2.5 of the Supplementary Material gives the proof.

Efficiency is not required for asymptotic prediction coverage, as shown
in Section~\ref{sec:prediction-coverage}. Yet gains in calibration
accuracy, for example from using $\widehat F_{\mathrm{AIPW}}(q)$ in
place of $\widehat F_{\mathrm{IPW}}(q)$, can tighten the resulting
prediction regions under additional regularity conditions. Sections S8.1--S8.2 of the Supplementary Material quantify this relationship for a fixed trained score.

\subsection{Finite-sample and asymptotic prediction coverage}\label{sec:prediction-coverage}
Proposition~\ref{prop:coverage-efficiency} concerns the efficiency of
$\widehat F_{\mathrm{AIPW}}(q)$ at a fixed $q$. To establish a coverage
guarantee for the calibrated region, however, we need to control
calibration error uniformly over $q$.

Calibration error combines sampling variation and nuisance-model bias.
For the sampling term, we use concentration inequalities to bound the
expected maximum fluctuation across thresholds and add a margin so that
the resulting bound holds with probability at least $1-\delta$.
Theorem~\ref{thm:identity} supplies the bias bound $c^{-1}r_Cr_Q$.

\begin{theorem}[Finite-sample coverage]\label{thm:split}
Condition on training, so that the fitted $(\widehat G,\widehat m)$ and their errors $(r_C,r_Q)$ are fixed. For $0<\delta<1$, let
\[
 e_n(\delta,c)=\left(\frac2c-1\right)
 \left\{\left(\frac{\pi}{2n}\right)^{1/2}
       +\left(\frac{\log(1/\delta)}{2n}\right)^{1/2}\right\}.
\]
Under the conditions of Theorem~\ref{thm:identity}, with probability at least $1-\delta$ over calibration,
\[
 \sup_q|\widehat F_{\mathrm{AIPW}}(q)-F(q)|
 \le e_n(\delta,c)+c^{-1}r_Cr_Q.
\]
Let $\Gamma_\tau^*$ and $Z^*$ denote the target and covariates for an independent new patient drawn from the same population as the calibration patients. The prediction region in~\eqref{eq:cutoff} satisfies, with probability at least $1-\delta$ over the calibration sample $\calD$,
\begin{equation}\label{eq:coverage}
 \pr\{\Gamma_\tau^*\in\widehat{\calC}(Z^*)\mid\calD\}
 \ge1-\alpha-e_n(\delta,c)-c^{-1}r_Cr_Q.
\end{equation}
\end{theorem}

Conditional on training, the calibration sample produces a random estimated coverage curve
$\widehat F_{\mathrm{AIPW}}(q)$ and threshold $\widehat q$. For at least a $1-\delta$ fraction of calibration samples, the resulting region satisfies~\eqref{eq:coverage}, with coverage averaged over new patients across covariate values. Averaging also over calibration samples gives the usual marginal guarantee, with an additional error allowance of at most $\delta$.

Increasing the calibration sample reduces the sampling variation term $e_n(\delta,c)$, while improving either nuisance fit reduces the bias term $c^{-1}r_Cr_Q$. The bound parallels the univariate survival guarantee of \citet{farina2026}. Sections S3--S4 of the Supplementary Material derive the uniform error bound for the estimated coverage curve and show how it yields the prediction coverage guarantee.

Asymptotic coverage follows when both error terms vanish. For fixed $c$
and $\delta$, the sampling term $e_n(\delta,c)$ tends to zero at rate
$n^{-1/2}$ as $n\to\infty$. The nuisance bias term $c^{-1}r_Cr_Q$
vanishes if either fit is consistent and the other error remains bounded.
\begin{corollary}[Asymptotic coverage]\label{cor:asymptotic-coverage}
Under Theorem~\ref{thm:split}, suppose $c>0$ is fixed, $n\to\infty$ and $r_Cr_Q=o_p(1)$. Then
\begin{equation*}
 \liminf_{n\to\infty}\pr\{\Gamma_\tau^*\in\widehat{\calC}(Z^*)\}\ge1-\alpha.
\end{equation*}
\end{corollary}

Finally, the component-specific procedure in Section~\ref{sec:implementation} assigns separate error levels to the targets. Applying Theorem~\ref{thm:split} to each component yields the following component-specific and joint coverage guarantees.
\begin{proposition}[Component-specific error allocation]\label{prop:unequal-allocation}
Consider the componentwise region $\widehat{\calC}_{\rm alloc}$ constructed in \eqref{eq:componentwise_region}. Let $\alpha_j>0$ with $\sum_{j=1}^d\alpha_j\le\alpha$ be chosen in advance or using only independent training data.
Suppose each component satisfies Theorem~\ref{thm:split} with common lower bound $c$. Let $r_{C,j}$ and $r_{Q,j}$ be the errors in Section~\ref{sec:bias-efficiency}, evaluated for its scalar target. Conditional on training, with probability at least $1-\delta$ over calibration, the component-specific prediction sets satisfy
\[
 \pr\{\Gamma_{\mathcal A_j}^*(\tau)\in I_j(Z^*)\mid\calD\}
 \ge1-\alpha_j-e_n(\delta/d,c)-c^{-1}r_{C,j}r_{Q,j},
 \qquad j=1,\ldots,d.
\]
Consequently, the joint prediction region satisfies
\[
 \pr\{\Gamma_\tau^*\in\widehat{\calC}_{\rm alloc}(Z^*)\mid\calD\}
 \ge1-\alpha-d\,e_n(\delta/d,c)
       -c^{-1}\sum_{j=1}^d r_{C,j}r_{Q,j}.
\]
For fixed $d$ and $c>0$, if $r_{C,j}r_{Q,j}=o_p(1)$ for every component as $n\to\infty$, then
$\liminf_{n\to\infty}\pr\{\Gamma_\tau^*\in\widehat{\calC}_{\rm alloc}(Z^*)\}\ge1-\alpha$.
\end{proposition}
Section S7.4 of the Supplementary Material gives the proof.

\section{Simulation studies}\label{sec:simulations}
\subsection{Design}
We first considered recurrent hospitalisation as in Fig.~\ref{fig:state-models}(a). Patients began at home and could be admitted and discharged repeatedly before death. The baseline predictors $Z=(Z_{\cdot 1},Z_{\cdot 2})^{\mathrm T}$ comprised a continuous measure of severity and an independent binary risk factor. We drew $Z_{\cdot 1}$ from a standard normal distribution truncated to $[-2.5,2.5]$ and $Z_{\cdot 2}$ from a Bernoulli distribution with probability $1/2$. Higher values of either covariate increased the rates of admission and death. Greater severity also slowed discharge, and a previous admission increased the readmission rate.

We chose rates to reflect hospitalisation patterns in heart failure trials such as HF-ACTION, a randomised trial of exercise training in chronic systolic heart failure \citep{oconnor2009hfaction}. The simulated admission rate and mean length of stay were close to those reported under usual care \citep{mao2023whilealive,reed2010hfaction}. Follow-up ended at death, censoring or $\tau=1$ year, whichever came first. Censoring was independent of the disease history given $Z$ and more frequent at higher values of either covariate. We set the censoring rate to give 20, 40 or 60\% censoring before death or one year. The Supplementary Material gives the transition and censoring rates and reports a separate simulation study of progressive disease.

\subsection{Prediction of days alive and days alive out of hospital}
For a new patient with covariates $Z$, we predicted days alive,
$\Gamma_{\{0,1\}}(\tau)=D\wedge\tau$, and days alive out of hospital,
$\Gamma_{\{0\}}(\tau)$, with their standardised scores joined as
in~\eqref{eq:maximum_joint} using $a_1=a_2=1$.
We fitted transition-specific Cox models to the censored histories of 500 training patients. For patients at home at the same time with the same $Z$, the working Markov model used a common admission rate regardless of prior admissions. Its predicted means and standard deviations for the two outcomes gave the joint score in Section~\ref{sec:implementation}. A separate Cox regression on $Z$ supplied the censoring weights.

We held the trained score fixed and calibrated it for 90\% joint coverage
in an independent sample of $n=200$, $400$, $600$ or $1{,}000$ patients.
Calibration using the same patients' complete, uncensored targets served as the benchmark.
For censored histories, we compared standard inverse probability weighting,
which evaluates the fitted weights at $D\wedge\tau$, with
$\widehat F_{\mathrm{IPW}}$ from Section~\ref{sec:calibration}, which
evaluates them at $\mathcal T_q$.
For $\widehat F_{\mathrm{AIPW}}$ in~\eqref{eq:aipw}, we used the fitted
transition models to estimate the probability that both outcomes would fall
in the candidate region, given $Z$, the current state and the occupation
times accrued so far. Each method selected its cutoff by~\eqref{eq:cutoff}.
In each repetition, we evaluated joint coverage in $2{,}000$ new patients
and averaged the interval width for each outcome over those patients.

\begin{figure}[!htbp]
\centering

\includegraphics{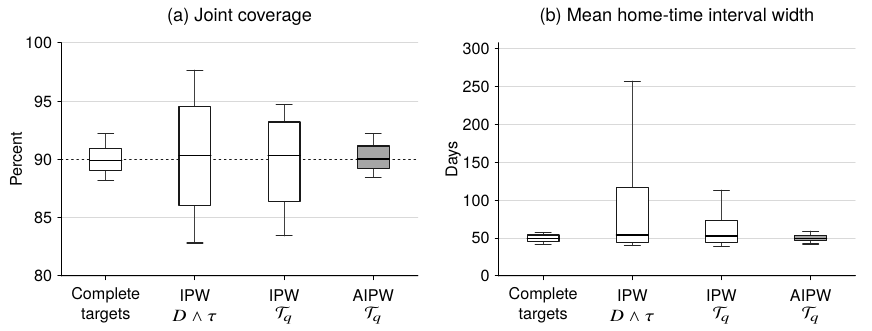}
\caption{Joint coverage and home-time width at 40\% censoring. Whiskers span the 10th to 90th percentiles.}
\label{fig:hf-stability}
\end{figure}

For calibration sample size $n=400$,
Table~\ref{tab:hf-calibration} reports mean joint coverage and the mean,
median and 90th percentile of these average widths at each censoring level.
All four methods gave mean joint coverage close to 90\%. Evaluating the
weights at $\mathcal T_q$ instead of $D\wedge\tau$ reduced mean interval
widths for both outcomes, with the largest reduction at 60\% censoring.
Augmentation further reduced mean and median widths, bringing them close
to the complete-target benchmark at each censoring level. Both steps also
reduced variation across repetitions. As Fig.~\ref{fig:hf-stability} shows
at 40\% censoring, very large mean widths for days alive out of hospital
occurred less often with weighting at $\mathcal T_q$ and were rarer still
with augmentation. Joint coverage also varied less with augmentation.
Joint coverage and interval widths showed similar patterns at $n=200$,
$600$ and $1{,}000$, as reported in the Supplementary Material.

\begin{table}[!htbp]
\caption{Joint coverage and interval widths for recurrent hospitalisation}\label{tab:hf-calibration}
\centering\small
\setlength{\tabcolsep}{3pt}
\begin{tabular*}{\textwidth}{@{\extracolsep{\fill}}lrrrrrrr@{}}
\toprule
 & Joint & \multicolumn{3}{c}{Days alive} & \multicolumn{3}{c}{Days alive out of hospital} \\
\cmidrule(lr){3-5}\cmidrule(l){6-8}
Calibration & coverage (\%) & Mean & Median & 90th & Mean & Median & 90th \\
\midrule
\addlinespace[4pt]\multicolumn{8}{@{}l}{\textit{20\% censoring}} \\
Complete targets & 90.0 & 46.0 & 45.6 & 54.8 & 50.5 & 50.1 & 59.1 \\
IPW at $D\wedge\tau$ & 90.0 & 62.5 & 47.2 & 93.1 & 66.9 & 51.4 & 97.7 \\
IPW at $\mathcal T_q$ & 89.8 & 53.1 & 46.8 & 66.6 & 57.6 & 51.2 & 71.1 \\
AIPW at $\mathcal T_q$ & 90.2 & 46.6 & 46.1 & 55.6 & 51.1 & 50.6 & 60.1 \\[2pt]
\addlinespace[2pt]\multicolumn{8}{@{}l}{\textit{40\% censoring}} \\
Complete targets & 90.0 & 45.3 & 45.2 & 53.0 & 49.9 & 49.6 & 56.8 \\
IPW at $D\wedge\tau$ & 90.2 & 95.3 & 50.0 & 251.3 & 99.5 & 54.2 & 256.9 \\
IPW at $\mathcal T_q$ & 89.5 & 63.6 & 48.2 & 108.5 & 68.0 & 52.7 & 113.0 \\
AIPW at $\mathcal T_q$ & 90.2 & 46.1 & 45.1 & 54.2 & 50.7 & 49.6 & 58.9 \\[2pt]
\addlinespace[2pt]\multicolumn{8}{@{}l}{\textit{60\% censoring}} \\
Complete targets & 89.9 & 45.7 & 45.8 & 54.5 & 50.3 & 50.4 & 59.6 \\
IPW at $D\wedge\tau$ & 89.9 & 124.7 & 48.2 & 365.0 & 128.2 & 52.6 & 365.0 \\
IPW at $\mathcal T_q$ & 88.8 & 70.6 & 48.6 & 149.0 & 75.0 & 53.0 & 154.2 \\
AIPW at $\mathcal T_q$ & 89.8 & 46.9 & 46.5 & 57.4 & 51.5 & 50.8 & 62.0 \\
\bottomrule\end{tabular*}
\end{table}

\subsection{Joint coverage under model misspecification}

To study the effect of omitting a prognostic covariate from the nuisance models, we made the binary risk indicator $Z_{\cdot 2}$ predict early loss to follow-up as well as subsequent hospitalisation and death. High-risk patients were therefore underrepresented later in follow-up, which could lead to bias under a faulty
censoring model. We considered two censoring models, both including the continuous covariate $Z_{\cdot 1}$. The correctly specified model also included $Z_{\cdot 2}$, whereas the misspecified model omitted it. We paired each censoring model with two specifications of the transition models for augmentation. The correctly specified augmentation model included both covariates, whereas the misspecified model omitted $Z_{\cdot 2}$.

We set up 500 repetitions. In each repetition, we used 300 training patients to fit the joint score in~\eqref{eq:maximum_joint} for both IPW and AIPW. On the same independent sample of 400 calibration patients, each method selected $\widehat q$ by~\eqref{eq:cutoff} to target 90\% joint coverage. We estimated joint coverage in 5000 new patients as the proportion whose prediction region covered both days alive and days alive out of hospital.

\begin{figure}[!htbp]
\centering
\includegraphics{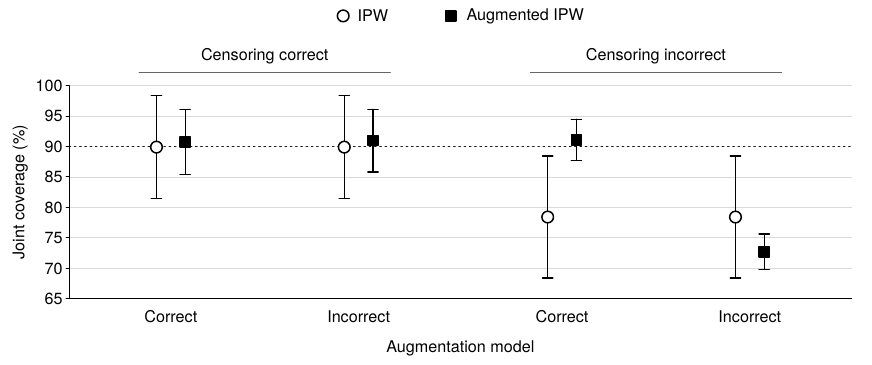}
\caption{Joint coverage after calibration on $n=400$ patients at $1-\alpha=90\%$
over 500 repetitions. Points show means and bars extend one standard deviation above and below each mean. }\label{fig:hf-dr}
\end{figure}

When both nuisance models were correctly specified, IPW and AIPW both attained mean joint coverage close to 90\%, as shown in Fig.~\ref{fig:hf-dr}. With the censoring model correctly specified, AIPW retained near-nominal coverage when the augmentation model omitted $Z_{\cdot 2}$. When the censoring model instead omitted $Z_{\cdot 2}$, IPW coverage fell to 78.4\%. With the same weights, AIPW attained 91.1\% coverage when the augmentation model was correctly specified, showing double robustness. 
Finally, when both models omitted $Z_{\cdot 2}$, AIPW coverage fell to 72.7\%. The Supplementary Material gives the widths and paired comparisons.

\subsection{Additional numerical experiments}

We also considered the progressive disease model in
Fig.~\ref{fig:state-models}(b), restricted to remission, relapse and
death. For patients initially in remission, we predicted days alive and
days in remission over one year. We used Cox, forest and neural network
transition models for augmentation, with censoring ranging from 20\% to
60\%. At 60\% censoring, augmentation improved joint coverage from
86.2\% with IPW to the 90\% target. All three transition models gave near-nominal coverage at each censoring level.
Differences in mean interval width among the four calibration methods were smaller here than in the recurrent hospitalisation study, as shown in Table~\ref{tab:hf-calibration} and Supplementary Table~S9. This may
reflect the simpler history available for predicting coverage, with at
most one relapse before death compared with a sequence of admissions
and discharges.
Supplementary Tables~S9--S11 give the calibration and model comparisons as well as some sensitivity analysis.

\section{Clinical applications}\label{sec:applications}
\subsection{Days in and out of hospital in HF-ACTION}\label{sec:clinical-application}
The HF-ACTION trial evaluated exercise training plus usual care versus usual care alone in 2,331 patients with symptomatic systolic heart failure. Its primary outcome was all-cause death or hospitalisation \citep{oconnor2009hfaction}. The available data include 2,130 participants, of whom 2,127 have follow-up beyond randomisation for both hospitalisation and death.

We jointly predict days alive out of hospital, $\Gamma_{\{0\}}(\tau)$, and days in hospital, $\Gamma_{\{1\}}(\tau)$, over the $\tau=365$ days after randomisation. We reconstruct admissions and discharges through death, the end of follow-up for either outcome or day 365, whichever comes first. Within one year, 99 patients die, and censoring leaves the joint outcome unknown for 79 patients, or 3.7\% of the cohort.

We randomly divide the patients into four nearly equal groups and use each group in turn for evaluation. In each rotation, the next group in cyclic order serves for calibration and the remaining two for training. Ten such random partitions give 40 splits and ten held-out predictions per patient.
As in the recurrent-hospitalisation simulations, we train the score using a Markov model with separate Cox models for admission and discharge. The Cox model for death includes current hospitalisation status as a covariate. All three models use time since randomisation and the same 14 baseline predictors. These are age, sex, body mass index, treatment assignment, New York Heart Association class, left ventricular ejection fraction, systolic blood pressure, blood urea nitrogen, history of diabetes, number of heart-failure admissions in the previous six months, exercise-test duration and modality, and the physical limitation and symptom stability scores from the Kansas City Cardiomyopathy Questionnaire. 
In addition to joint prediction of $\Gamma_{\{0\}}(\tau)$ and $\Gamma_{\{1\}}(\tau)$,
we also calibrate a separate interval for the home time $\Gamma_{\{0\}}(\tau)$.

As in the simulations, we compare IPW at death or the horizon, IPW at the coverage determining time and AIPW at the coverage determining time. A separate Cox model with the same baseline predictors supplies the censoring weights. For augmentation, we refit the transition models and simulate continuations from the patient's current state, retaining the time already spent in and out of hospital. Each method uses~\eqref{eq:cutoff} to select the smallest cutoff at which estimated calibration coverage reaches the 90\% target. We evaluate all three methods on the same held-out patients using the augmented coverage estimator.

As shown in Table~\ref{tab:hfa-coverage-width}, all three methods appear well calibrated, with estimated coverage close
to the nominal 90\% for both the joint region and the home-time interval.
There is also little difference among
the methods in the mean width of the home-time interval or in how much
that mean varies across splits. With little censoring by one year, most
calibration targets are observed, leaving little scope for weighting or
augmentation to affect the results.
For additional analyses, Section~S10.15 of the Supplementary Material provides the coverage estimates by symptom class, bootstrap intervals for overall coverage and the projected widths of the joint regions.

\begin{table}[!htbp]
\caption{One-year coverage and home-time interval width across 40 splits in HF-ACTION}\label{tab:hfa-coverage-width}
\centering\small
\setlength{\tabcolsep}{3pt}
\begin{tabular*}{\textwidth}{@{\extracolsep{\fill}}lrrrrrr@{}}
\toprule
 & \multicolumn{2}{c}{Joint coverage} & \multicolumn{2}{c}{Home-time coverage} & \multicolumn{2}{c}{Mean home-time width} \\
 & \multicolumn{2}{c}{(\%)} & \multicolumn{2}{c}{(\%)} & \multicolumn{2}{c}{(days)} \\
\cmidrule(lr){2-3}\cmidrule(lr){4-5}\cmidrule(l){6-7}
Calibration & Mean & SD & Mean & SD & Mean & SD \\
\midrule
IPW at $D\wedge\tau$ & 89.8 & 2.2 & 90.0 & 2.1 & 28.9 & 3.1 \\
IPW at $\mathcal T_q$ & 89.9 & 1.9 & 89.9 & 2.1 & 28.8 & 3.0 \\
AIPW at $\mathcal T_q$ & 90.4 & 1.7 & 90.4 & 1.8 & 29.3 & 2.9 \\
\bottomrule\end{tabular*}
\end{table}

\FloatBarrier

\subsection{Relapse-free survival in node-positive breast cancer}\label{sec:gbsg}
The German Breast Cancer Study Group cohort included 720 women with node-positive breast cancer. The randomised trial compared three versus six cycles of chemotherapy, with or without tamoxifen, in 473 of these women. The trial assessed relapse-free and overall survival \citep{schumacher1994gbsg}. Our analysis uses the 686 women with complete baseline data \citep{royston2013external}. 

We consider the illness--death model obtained by omitting the metastasis state from Fig.~\ref{fig:state-models}(b). A woman starts in remission and may die before or after relapse. At study entry, we seek a 90\% prediction interval for $\Gamma_{\{0\}}(\tau)$, the time she spends in remission over the next $\tau=5$ years. For 40.5\% of the women, follow-up ends before five years while they are still in remission. The baseline predictors are age, menopausal status, tumour size, histological grade, number of positive lymph nodes, progesterone and oestrogen receptor levels and hormonal treatment.

We rotate the training, calibration and evaluation groups as in HF-ACTION. A Cox model for recurrence or death gives the mean and standard deviation of restricted relapse-free time given the baseline predictors. The score measures the absolute deviation from the predicted mean in units of the predicted standard deviation. A second Cox model with the same predictors provides the censoring weights. For augmentation, we use the event model to estimate the chance that a woman's restricted relapse-free time falls within her interval, given how long she has already lived without recurrence.

\begin{figure}[!htbp]
\centering

\includegraphics{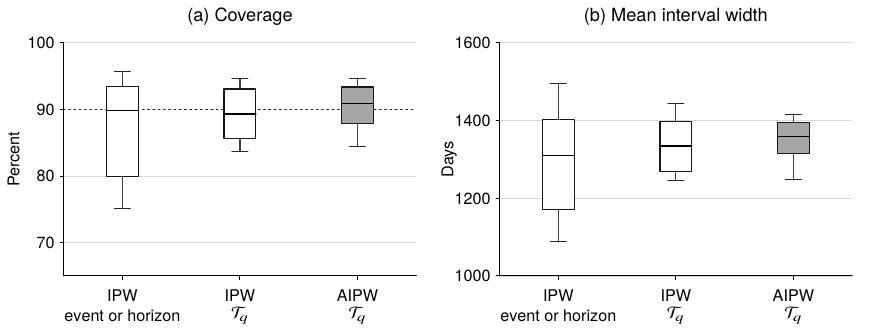}
\caption{Box plots of coverage and mean interval width for five-year relapse-free time in the German Breast Cancer Study Group cohort.}
\label{fig:gbsg-performance}
\end{figure}

With heavier censoring than in HF-ACTION, the choice of calibration method makes a larger difference. For IPW at the first event or the horizon, estimated coverage averages 86.8\% and varies widely across splits, as Fig.~\ref{fig:gbsg-performance} shows. Using the coverage determining time reduces this variation and brings mean coverage closer to the 90\% target for both IPW and AIPW, at 89.2\% and 90.3\%, respectively. Mean interval width also varies less across splits for the two determining-time methods, with the least variation for AIPW.
To illustrate individual predictions, Fig.~S4 in the Supplementary Material shows the AIPW intervals from all 40 splits for two women with different covariate profiles.

\section{Discussion}\label{sec:discussion}
We have developed a conformal prediction approach for summaries of a multistate
process, such as state occupation times.  Prediction is restricted to a clinically
meaningful horizon, avoiding the non-identifiability
of the target's upper tails due to right censoring
\citep{candes2023,farina2026}. Within the restricted window,
we have introduced the coverage determining time to make fuller use of
partial histories in calibration. Combining inverse weighting of these decisions with coverage
predictions for unresolved histories has yielded a doubly robust estimator that
can attain the efficiency bound for a fixed region's coverage. Both
theoretical and empirical results support the efficiency and stability
of the proposed calibration procedure, particularly when censoring is substantial.

Our numerical studies emphasise state occupation times. However, the same
calibration framework applies to general functionals of the outcome trajectory.
For example, when the state labels increase with severity,
$\Gamma(\tau)=\max_{0\le t\le\tau}Y(t)$ records the most severe state reached
by $\tau$, as defined in Section~\ref{sec:targets}. For this categorical target, the only change from the procedure for
quantitative occupation times is the score function. We score each
candidate state $y$ by $s(z,y)=1-\widehat{\pr}\{\Gamma(\tau)=y\mid Z=z\}$,
calibrate the cutoff $\widehat q$ as before, and retain all states with
$s(z,y)\le\widehat q$. One could also calibrate separate cutoffs within each category, for example,
to limit how often the set omits death among patients who die
by $\tau$.

A prediction region calibrated in one population may have a different behaviour
in another if the distributions of baseline characteristics differ. For example,
patients seen in a heart failure clinic may differ from the calibration
sample of HF-ACTION participants
 in age and baseline severity.
Such a covariate shift violates the assumption that the calibration
patients and the new patient are exchangeable.
However, if the trial includes the baseline profiles seen in the clinic and prognosis
within each profile is the same in both settings, trial participants can be
weighted to reflect the clinic's case mix \citep{tibshirani2019shift,qiu2023}.
An extension of our approach would combine these weights with the adjustment
for censoring to select a prediction cutoff for the clinic population.

We have focused on split conformal procedures, with separate training
and calibration samples. Full conformal prediction uses all patients
for both steps. With complete outcomes, one adds each candidate outcome
for the new patient to the data in turn, refits the score and ranks the
new patient's score among all patients' scores \citep{lei2018}. 
Under censoring, some score comparisons may remain unresolved, leaving
the new patient's rank only partially determined. One approach is to
weight each observed comparison by the inverse probability of observing
it. However, refitting a multistate model and recomputing the weighted
comparisons across all patient histories for each candidate outcome
would be computationally demanding. It would be of interest to explore
whether these calculations can be simplified for selected trajectory
functionals while retaining a coverage guarantee for the resulting
full conformal intervals.

We have considered prediction at study entry. Dynamic prediction would allow us
to update a patient's prediction as new information comes in. At the
six-month visit in HF-ACTION, for example, a patient's admissions and time spent
in hospital to date could inform a prediction interval for days alive and out of
hospital over the next six months. The landmarking approach
\citep{vanhouwelingen2007landmark} could be adapted to calibrate this interval using patients alive and under observation at six months. 
If predictions
are updated at successive visits, a further question is how to achieve
simultaneous coverage for the intervals given to the same patient \citep{tassopoulou2025}.

\section*{Supplementary material}
The Supplementary Material includes proofs, the survival-time specialisation, secondary calibration and region-size results, and computational details for the simulations and both clinical applications.

\section*{Data availability}
HF-ACTION data are available by application to the NHLBI Biologic Specimen and Data Repository Information Coordinating Center (BioLINCC), accession HLB01261520a. Records from the German Breast Cancer Study Group \citeyearpar{gbsg2024data} are available as \texttt{gbsg} in the R package \texttt{survival}.

\bibliographystyle{abbrvnat}
\bibliography{references}
\end{document}